\documentclass[twocolumn,showpacs,preprintnumbers,nofootinbib]{revtex4-1}%
\usepackage{amsfonts}
\usepackage{amsmath}
\usepackage{graphicx}
\usepackage{dcolumn}
\usepackage{bm}
\usepackage{slashed}
\usepackage{amssymb}
\usepackage{MnSymbol}
\usepackage{xcolor}
\usepackage{verbatim}
\usepackage{todonotes}
\usepackage{grffile}

\usepackage{soul}

\begin{document}
\title{Perturbative thermodynamics at nonzero isospin density for cold QCD}

%\title{The equation of state in pQCD for nonvanishing isospin density}
\author{Thorben Graf}
\affiliation{Institute for Theoretical Physics, Goethe University,
Max-von-Laue-Stra{\ss}e 1, D--60438 Frankfurt am Main, Germany}
\author{Juergen Schaffner-Bielich}
\affiliation{Institute for Theoretical Physics, Goethe University,
Max-von-Laue-Stra{\ss}e 1, D--60438 Frankfurt am Main, Germany}
\author{Eduardo S. Fraga}
\affiliation{Instituto de Fisica, Universidade Federal do Rio de Janeiro,
Caixa Postal 68528 Rio de Janeiro, RJ 21941-972, Brasil}

\begin{abstract}
  We use next-to-leading-order in perturbation theory to investigate the
  effects of a finite isospin density on the thermodynamics of cold strongly
  interacting matter. Our results include nonzero quark masses and are
  compared to lattice data.
\end{abstract}

\pacs{11.10.Wx,12.38.Bx,12.38.Mh,21.65.Qr}
\keywords{xxx}

\maketitle

%\date{\nodate}

%\listoftodos

%%%%%%%%%%%%%%%%%%%%%%%%%%%%%%%%%%%%%%%%%%%%%%%%%%%%%%%%%%%%%%%
\section{Introduction}

The thermodynamics of strongly interacting matter at non-vanishing isospin
chemical potential, $\mu_I$, is relevant in different realms of physics, since
there are several systems where the amounts of protons and neutrons are not
the same. In the formation process of neutron stars, the initial proton
fraction in supernovae is $\sim 0.4$, which reduces with time to values of
less than $0.1$ in cold neutron stars
\cite{Prakash:2000jr,Weber:2004kj}. In the early universe, shortly
after the Big Bang, a large asymmetry in the lepton sector that could shift
the equilibrium conditions at the cosmological quark-hadron transition is
allowed \cite{Schwarz:2009ii}. And, of course, in high-energy heavy-ion
collisions, the proton to neutron ratio is $\sim 2/3$ in Au or Pb beams.

The phase diagram of QCD at finite temperature and isospin density is rich in
phenomenology and has been investigated for over a decade
\cite{Son:2000xc,Son:2000by}. Since then, several studies were performed
within effective models, on the lattice and most recently even perturbatively
\cite{Toublan:2003tt,Kogut:2004zg,Kogut:2004qq,Toublan:2004ks,Kogut:2005qg,Sinclair:2006zm,Andersen:2006ys,deForcrand:2007uz,Cea:2012ev,
  Fraga:2008be,Palhares:2008px,Andersen:2007qv,Kamikado:2012bt,Sasaki:2010jz,Ueda:2013sia,Stiele:2013pma,Xia:2013caa,Kanazawa:2014lga,Endrodi:2014lja,Andersen:2015eoa}. Although
Monte Carlo simulations do not suffer from the sign problem since the fermion
determinant remains real at nonzero $\mu_I$, lattice calculations at non-zero
isospin have been performed so far with unphysical quark masses
\cite{Kogut:2004zg,Kogut:2004qq,Kogut:2005qg,Sinclair:2006zm,deForcrand:2007uz,Cea:2012ev,Endrodi:2014lja},
which still limits their quantitative predictive power.

In this paper, we use next-to-leading-order in perturbation theory to
investigate the effects of a finite isospin density on the thermodynamics of
cold ($T=0$) strongly interacting matter which includes nonzero quark
masses. Whenever possible, our results are compared to lattice data from
Ref.~\cite{Detmold:2012wc}. The paper is organized as follows: in Section II
we present a brief discussion of the physical scenario and our setup; in
Section III we show and discuss our results for the thermodynamical quantities
computed; and section IV contains our final remarks.

%%%%%%%%%%%%%%%%%%%%%%%%%%%%%%%%%%%%%%%%%%%%%%%%%%%%%%%%%%%%%%%
\section{Physical scenario and setup}

\begin{center}
  \begin{figure}
      %\vspace{-2.5cm}
    \begin{minipage}[c]{8cm}
      \includegraphics[width=\textwidth]{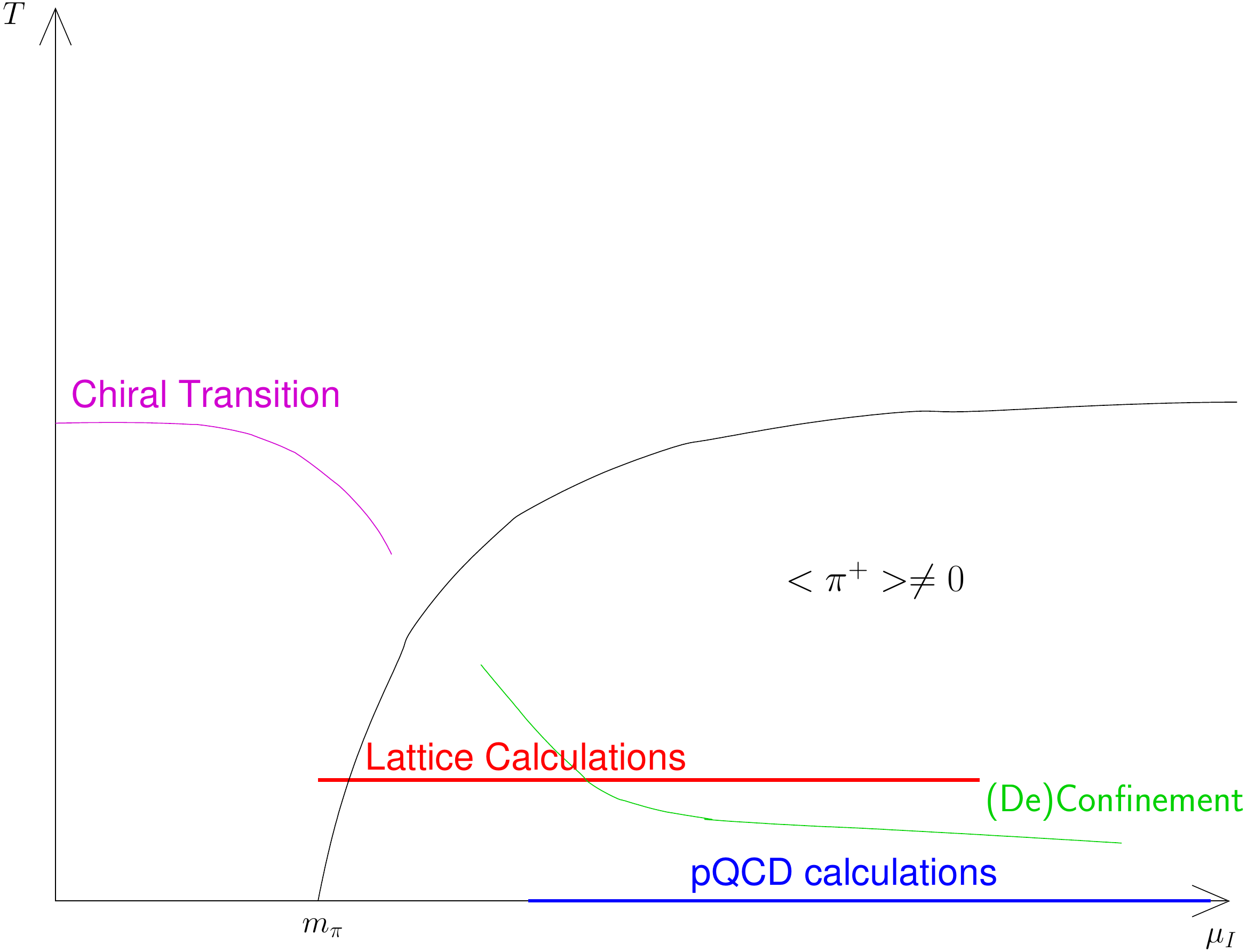}
    \end{minipage}
    %\vspace{-2.5cm}
    \caption{Cartoon of the phase diagram of QCD at finite temperature and isospin chemical potential based on results from Refs.~\cite{Cea:2012ev,Cohen:2015soa}.}
    \label{fig:1}
  \end{figure}
\end{center}

The phase diagram of QCD in the temperature versus isospin chemical potential
plane is illustrated in Fig.~\ref{fig:1}, which should be seen as a
cartoon. Along the temperature axis ($\mu_{I}=0$) there is no phase
transition, according to lattice calculations at physical quark masses
\cite{Aoki:2006we}. At high isospin density, for values of $\mu_I$ above the
pion mass $m_\pi$, pion condensation takes place for not too large
temperatures. At very high isospin density a Fermi liquid with Cooper pairing
is formed as a consequence of an attractive interaction between quarks in the
isospin channel \cite{Son:2000xc}. In contrast to the temperature versus
baryon chemical potential ($\mu_B$) plane, there is a first-order
deconfinement phase transition for large $\mu_I$ within the condensed phase,
as indicated by the green line in Fig.~\ref{fig:1}. The authors of
Ref.~\cite{Son:2000by} conjecture that the phase transition line ends at a
second-order point \footnote{Other investigations suggest different scenarios
  concerning the existence of this critical point
  \cite{Cohen:2015soa}}. According to Ref.~\cite{Cea:2012ev}, the chiral phase
transition is located along the purple line in Fig.~\ref{fig:1}.

%%
%\begin{center}
%  \begin{figure}
%    \centering
%    \vspace{5cm}
%    \begin{picture}(100,80)% width and height of the picture
%      \put(-100,0){\includegraphics[width=0.4\textwidth]{PDNew.pdf}}
%      \put(-120,195){$T$}
%      \put(-30,25){Lattice calculations}
%      \put(0,-5){pQCD calculations}
%      \put(130,30){(De)confinement transition}
%      \put(-90,145){Chiral transition}
%      \put(70,80){$<\pi^+>\neq0$}
%      \put(190,-5){$\mu_I$}
%      \put(-55,-5){$m_\pi$}
%    \end{picture}
%    \caption{Cartoon of the phase diagram of QCD at finite temperature and isospin chemical potential based on results from Refs.~\cite{Cea:2012ev,Cohen:2015soa}.}
%    \label{fig:1}
%  \end{figure}
%\end{center}
%%

%

Lattice calculations of Ref.~\cite{Detmold:2012wc} were run at nonzero $\mu_I$
and at a fixed temperature of $T=20$ MeV. The values of $\mu_I$ covered in the
simulations are indicated by the horizontal red line in Fig.~\ref{fig:1}. Our
perturbative calculations were performed for values of the isospin chemical
potential which are represented by the blue solid line, at $T=0$. This
difference should not be significant given the comparatively large values of
$\mu_{I}$, as was verified a posteriori.

The energy scale of the (de)confinement transition was computed in
Ref.~\cite{Cohen:2015soa} using an effective model description and found to be
quite low, $\Lambda_{\text{Con}}\approx15-50$ MeV. Numerical values for the
(de)confinement scale were also computed in
Refs.~\cite{Kogut:2004zg,deForcrand:2007uz,Cea:2012ev}.

The phenomenon of pairing mentioned above should not be relevant for our
perturbative study, in the same fashion that happens at nonzero (large) baryon
chemical potential. The gap $\Delta$ is exponentially suppressed for small
values of $g$, in the domain of validity of perturbation theory
\cite{Son:1998uk,Son:2000xc},
\begin{equation}
 \Delta=b|\mu_I|g^{-5}e^{-c/g} \quad,
\end{equation}
where $c=3\pi^2/2$ and $g=g(|\mu_I|)$ is the running coupling. We expect that
the corresponding gap for nonzero isospin chemical potential will stay below
$\Delta\sim300-400$ MeV and hence will give a subleading contribution to the
thermodynamic potential $\sim\mu_{I}^2\Delta^2$ \cite{Alford:2002kj}.

Since the lattice calculations (red line in Fig.~\ref{fig:1}) might cross the
deconfinement transition (green line) as conjectured in
Ref.~\cite{Cohen:2015soa}, one can expect that perturbative calculations could
provide a reasonable description of lattice results for large enough values of
$\mu_{I}$. With the help of Fig.~2 in Ref.~\cite{Cohen:2015soa}, a
quantitative statement about the scale of $\mu_I$ at which the deconfined
phase appears can be made: for $\mu_I\simeq4$ GeV the deconfinement phase
transition line crosses $T=20$ MeV, the value used in the lattice simulations
of Ref.~\cite{Detmold:2012wc} to which we compare our findings.

For $T=0$ the expressions for the thermodynamic potential are available in
analytic form up to $\mathcal{O}(\alpha_s^2)$. The one massive flavor
contribution (leading and next-to-leading order) in the $\overline{\text{MS}}$
scheme is given by (see,
e.g. Refs.~\cite{Fraga:2004gz,Kurkela:2009gj,Graf:2015tda})
\begin{eqnarray}
%  \begin{split}
  \label{eq-ThermPotTZero}
    \Omega^{(0)}&=&-\frac{N_C}{12\pi^2}\left[\mu u\left(\mu^2-\frac{5}{2}m^2\right)
    +\frac{3}{2}m^4\ln\left(\frac{\mu+u}{m}\right)\right], \\
    \label{eq-ThermPotTOne}
    \Omega^{(1)}&=&\frac{\alpha_sN_G}{16\pi^3}\left\{3\left[m^2\ln\left(\frac{\mu+u}{m}\right)-\mu u\right]^2-2u^4 \right. \nonumber \\
    &+& \left. m^2\left[6\ln\left(\frac{\Lambda}{m}\right)+4\right]\left[\mu u-m^2\ln\left(\frac{\mu+u}{m}\right)\right]\right\},
%  \end{split}
\end{eqnarray}
where $u\equiv\sqrt{\mu^2-m^2}$ and $N_C$ and $N_G$ are the numbers of colors and gluons, respectively. For calculations with 2+1 massive quark
flavors we introduce the isospin chemical potential in the following way:
\begin{equation}
  \label{eq-IsoIntro}
  \begin{split}
  \mu_I=\mu_u-\mu_d &,\quad \mu_q=\frac{\mu_u+\mu_d}{2}, \\
  \mu_u=\mu_q+\frac{1}{2}\mu_I &,\quad \mu_d=\mu_q-\frac{1}{2}\mu_I, \\
  \mu_s=0,
  \end{split}
\end{equation}
where $\mu_q$ is the quark chemical potential. We assume $\mu_q=0$ in what follows.

%%%%%%%%%%%%%%%%%%%%%%%%%%%%%%%%%%%%%%%%%%%%%%%%%%%%%%%%%%%%%%%
\section{Results}

In order to compare our results with those from lattice simulations presented
in Ref.~\cite{Detmold:2012wc}, we adjusted our parameters accordingly. The
strange quark chemical potential $\mu_s$ is chosen to be zero, and the vacuum
pion mass is taken to be $m_\pi=390$ MeV. This corresponds to light quark
masses $m_{u/d}$ and a strange quark mass $m_s$ given by
\begin{equation}
  m_{u/d}= 35 \text{ MeV and } m_s=875 \text{ MeV}\quad,
\end{equation}
as extrated from the GOR-relation \cite{GellMann:1968rz}. Since $\mu_s=0$, the
strange quark plays no role in our analysis.

Our calculations implement a running coupling $\alpha_s$
\cite{Vermaseren:1997fq,Eidelman:2004wy}
\begin{equation}
\alpha_s(\Lambda)=\frac{4\pi}{\beta_0L}\left[1-2\frac{\beta_1}{\beta_0^2}\frac{\ln L}{L}\right] \quad,
\label{eq-RunAlphas}
\end{equation}
where $L=2\ln(\Lambda/\Lambda_{\overline{\text{MS}}})$, $\beta_0=11-2N_f/3$
and $\beta_1=51-19N_f/3$. The scale $\Lambda_{\overline{\text{MS}}}$ and is
fixed by requiring $\alpha_s\simeq0.3$ at $\Lambda=2$ GeV
\cite{Eidelman:2004wy} and one obtains
$\Lambda_{\overline{\text{MS}}}\simeq380$ MeV. See also
Ref.~\cite{Fraga:2004gz} for details.  With these conventions, the only
freedom left is the choice of the renormalization scale $\Lambda$, which is set to $\Lambda=2\mu_I$ in all of our numerical simulations.

\begin{figure}
\includegraphics[width=\columnwidth]{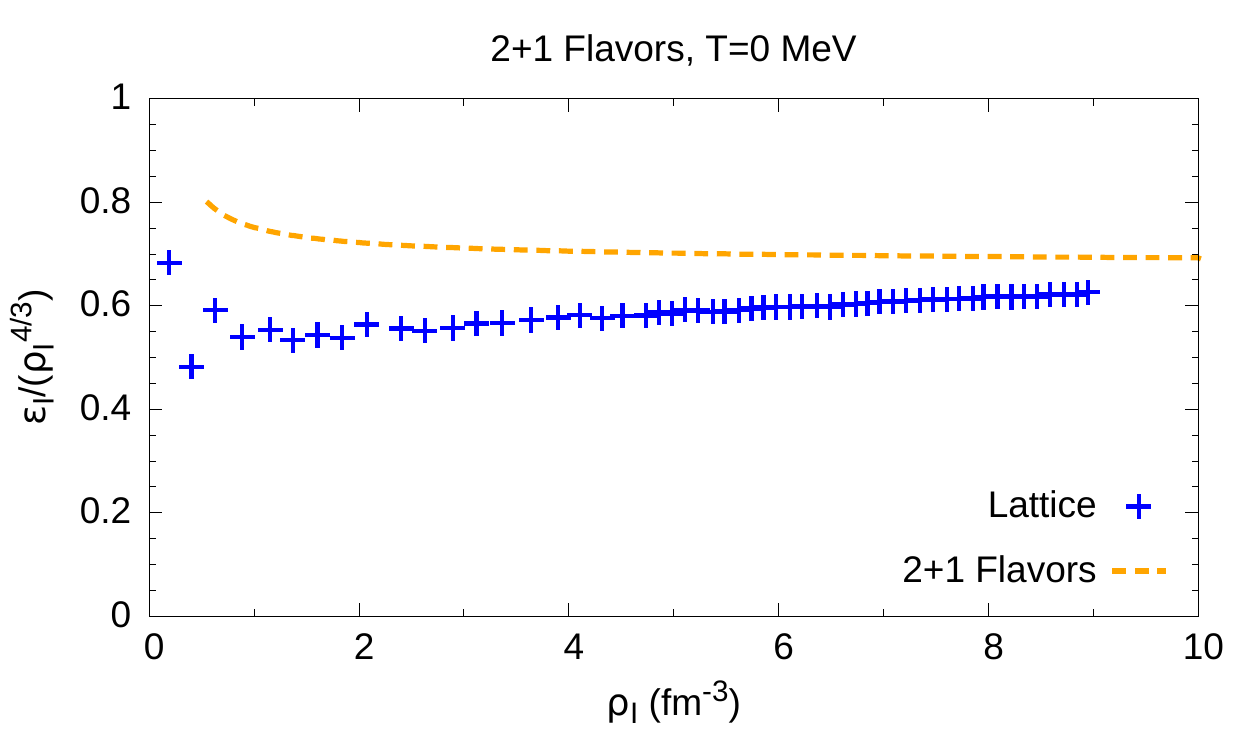}
\caption{Comparison of the ratio of energy density to (isospin
  density)$^{4/3}$ versus the
  isospin density with lattice results from Ref. \cite{Detmold:2012wc}.}
\label{fig:epsilon}
\end{figure}

\begin{figure}
\includegraphics[width=\columnwidth]{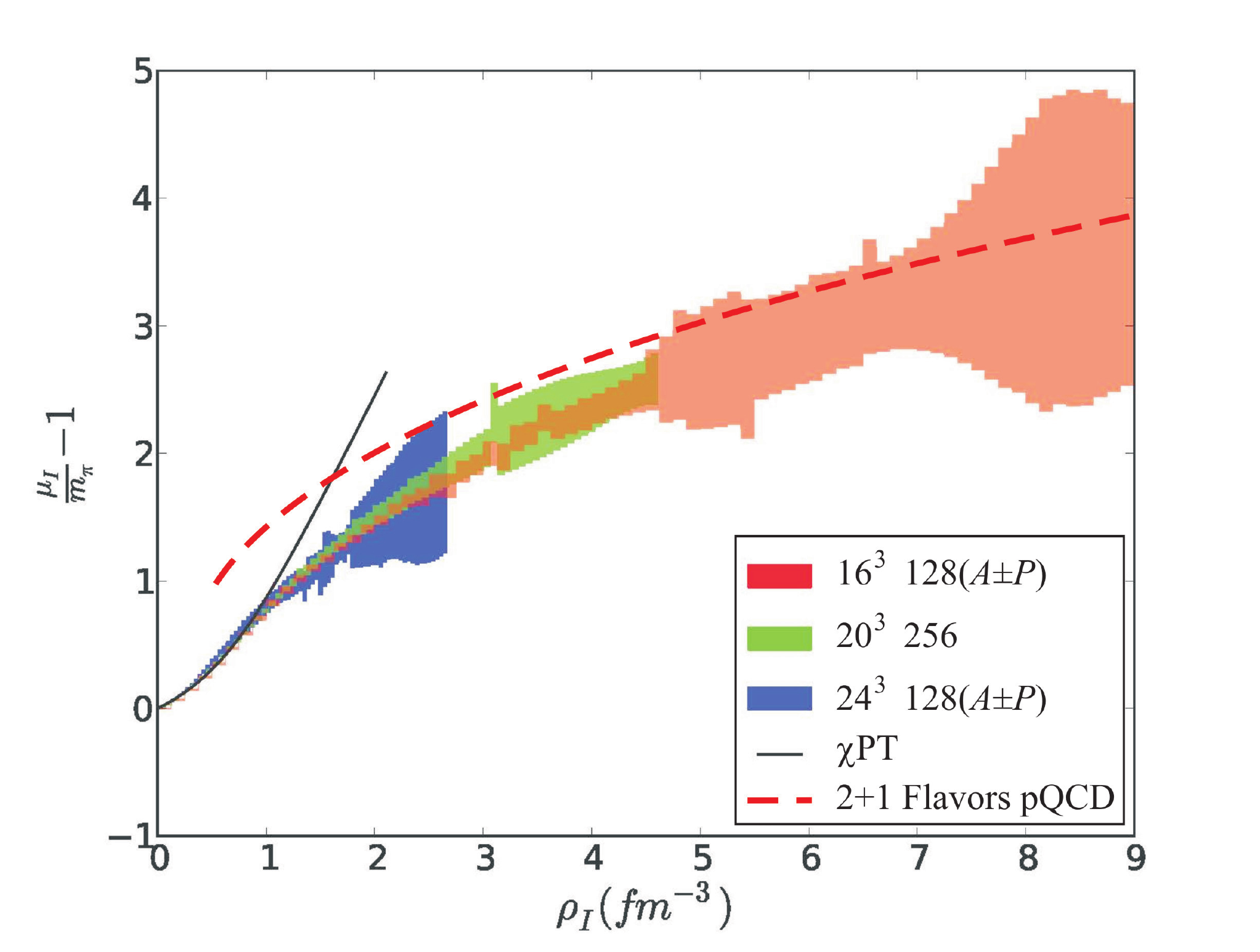}
\caption{Comparison of the isospin chemical potential versus the
  isospin density with lattice results from Ref.~\cite{Detmold:2012wc}.}
\label{fig:murho}
\end{figure}

From the thermodynamic potential, Eqns.~\eqref{eq-ThermPotTZero} and \eqref{eq-ThermPotTOne}, we have full access to all thermodynamical quantities, such as the pressure
\begin{equation}
  \Omega=-pV \quad,
\end{equation}
the isospin density $\rho_I$
\begin{equation}
  \rho_I=\frac{\partial p}{\partial \mu_I} \quad,
\end{equation}
and the energy density $\varepsilon$ (for $T=0$)
\begin{equation}
  \varepsilon=\frac{\partial p}{\partial \mu_I}\mu_I-p \quad.
\end{equation}
In Figs.~\ref{fig:epsilon} and \ref{fig:murho} we compare our results with
lattice data from Ref.~\cite{Detmold:2012wc}. In Fig.~\ref{fig:epsilon}, the
ratio of energy density to (isospin density)$^{4/3}$ is plotted against the
isospin density. One can see that for increasing isospin density the two
curves approach each other, as expected from asymptotic freedom, although
perturbation theory systematically overestimates this quantity within the
range of available lattice data extracted from Ref.~\cite{Detmold:2012wc}. We
stress, that the density dependence with a power of 4/3 is characteristic for
an ultrarelativistic Fermi gas, the asymptotic limit at high chemical
potentials. Note that an isospin density of roughly $9$ fm$^{-3}$ corresponds
to a value of $\mu_I=2$ GeV. In Fig.~\ref{fig:murho}, the isospin chemical
potential (subtracted by and normalized by the pion mass) is displayed versus
the isospin density. The results from pQCD agree well with those that
correspond to a band of lattice results extracted from
Ref.~\cite{Detmold:2012wc} for values of the isospin chemical potential larger
than about a few times the pion mass.

\begin{figure}
    \begin{minipage}[c]{9cm}
      \includegraphics[width=\textwidth]{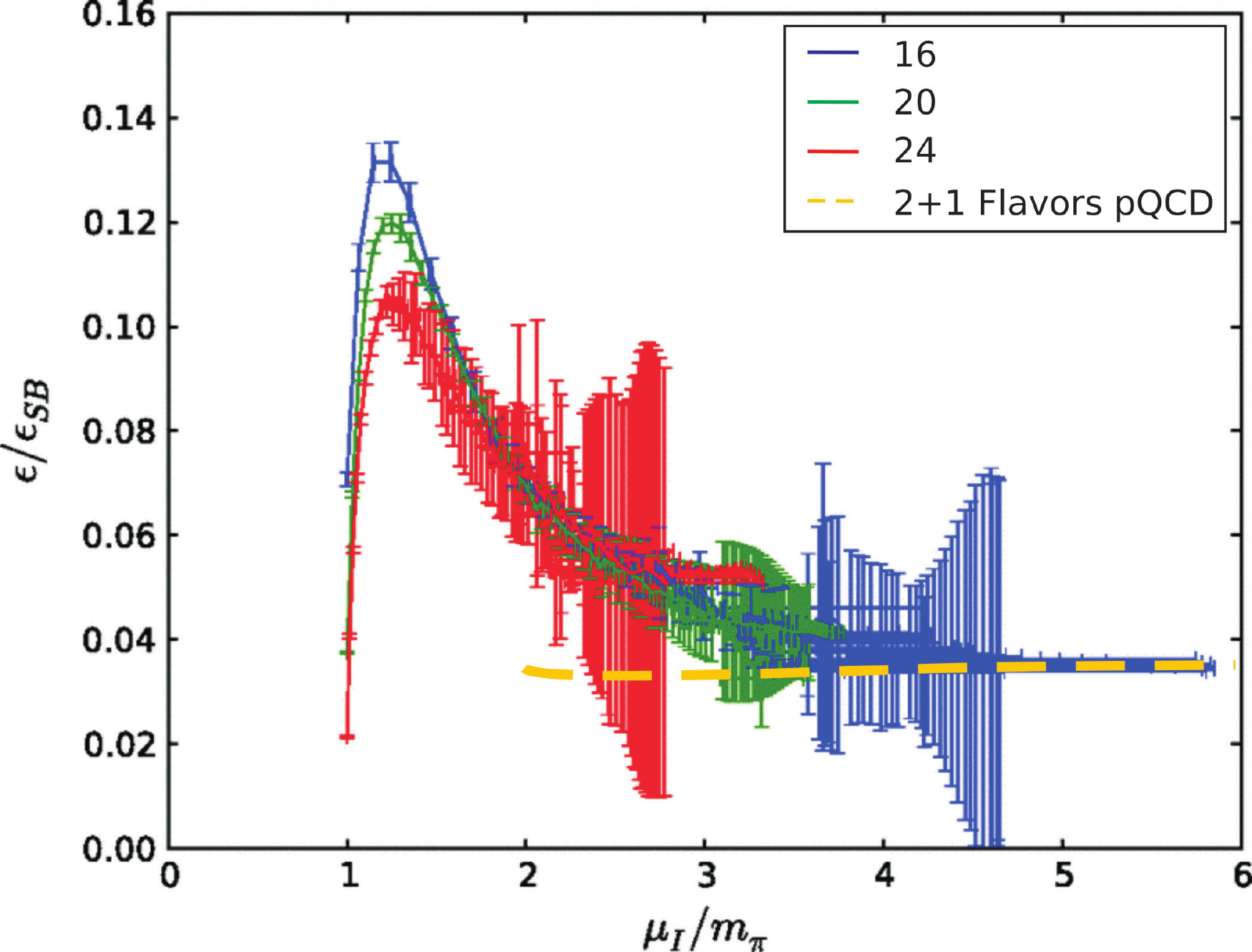}
    \end{minipage}
%    \begin{minipage}[c]{9cm}
%       \includegraphics[width=\textwidth]{EnergyDensity_TZero_Isotest_3.pdf}
%    \end{minipage}
    \caption{Energy density normalized by the isospin-related Stefan-Boltzmann
      (SB) form versus $\mu_{I}/m_{\pi}$. Lattice results from
      Ref.~\cite{Detmold:2012wc}.}
    \label{fig:3}
  \end{figure}

In Fig.~\ref{fig:3} we exhibit the energy density normalized by the
Stefan-Boltzmann (SB) form versus $\mu_{I}/m_{\pi}$, and also compare with the
corresponding band of lattice data extracted from
Ref.~\cite{Detmold:2012wc} who define the SB limit via the isospin chemical
potential as:
\begin{equation}
 \label{eq-Eps}
 \varepsilon_{SB}=\frac{N_fN_c}{4\pi^2}\mu_I^4 \quad.
\end{equation}
In terms of quark degrees of freedom the SB limit is given as a function of the
quark chemical potential
\begin{equation}
 \label{eq-EpsIso}
 \varepsilon_{SB}=\frac{N_fN_c}{4\pi^2}\mu^4=\frac{N_fN_c}{4\pi^2}\frac{\mu_I^4}{16} \quad.
\end{equation}
which gives via the relation $\mu=\frac{1}{2}\mu_I$ a factor 16 difference in
the corresponding SB limits. The latter one would be the limit for a gas of
quarks at zero temperature and high chemical potentials and hence also the SB
limit for pQCD calculations.

\begin{center}
  \begin{figure}
   \begin{minipage}[c]{9cm}
      \includegraphics[width=\textwidth]{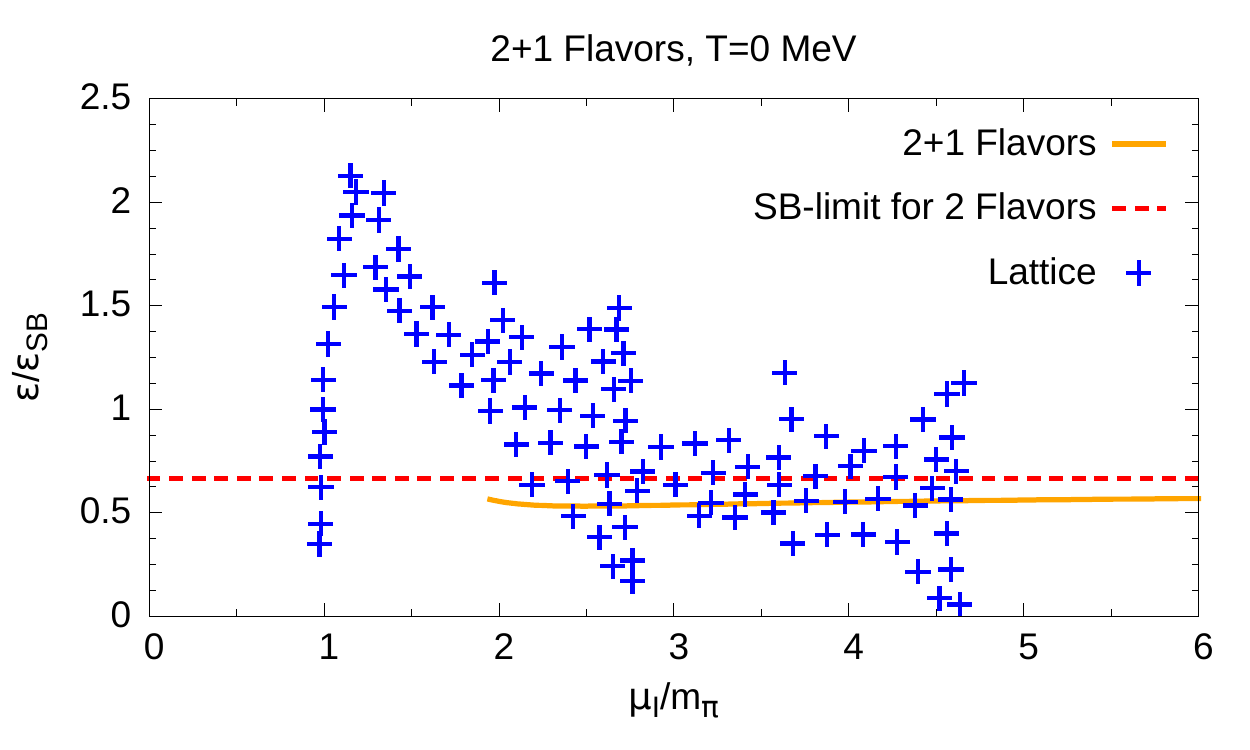}
   \end{minipage}
   \caption{Rescaled Energy density by an isospin-related factor 16 (see text
     for details) versus $\mu_{I}/m_{\pi}$. The pQCD results approach the SB
     limit related to the quarkchemical potential for two flavours as the
     strange quark is not appearing in the calculation.}
    \label{fig:4}
  \end{figure}
\end{center}

One sees in Fig.~\ref{fig:3} that for $\mu_I>2m_\pi$ the pQCD results are
compatible with the ones from the lattice. The peak at $\mu_I\approx m_\pi$
can not be reproduced since it is caused by the pion condensate which is not 
captured by perturbation theory. Simulations that are based on chiral
perturbation theory ($\chi$PT) are indeed able to calculate this maximum \cite{Carignano:2016rvs}.
By maximizing the static chiral Lagrangian density the authors derive an analytic
expression for the normalized energy density at the peak at leading-order. In general,
lattice data is well reproduced by $\chi$PT at leading-order for low densities, $\mu_I<2m_\pi$.
However, for $\mu_I > 2m_\pi$ the results of chiral perturbation theory 
asymptotically approaches zero as only pion degrees of freedom are 
incorporated. This is in contrast to the lattice data which reaches at
asymptotically high isospin chemical potentials our results from pQCD which
is based on quark degrees of freedom. In Fig.~\ref{fig:4} the same data of Fig.~\ref{fig:3}
is shown with regard to the SB limit for a gas of quarks,
i.e.\ rescaled by a factor of 16 which appears when $\mu=\frac{1}{2}\mu_I$.
The SB limit for 2 flavors (horizontal  line) is also sketched in Fig.~\ref{fig:4}
because the strange quark is not contributing in our calculations as $\mu_s=0$ so that
this SB limit should be considered as the actual limit of a free gas of quarks
and gluons. In this sense, our results (orange line) are obviously very close
to this limit which is consistent with the notion of asymptotic freedom.

\begin{center}
  \begin{figure}
    \begin{minipage}[c]{9cm}
      \includegraphics[width=\textwidth]{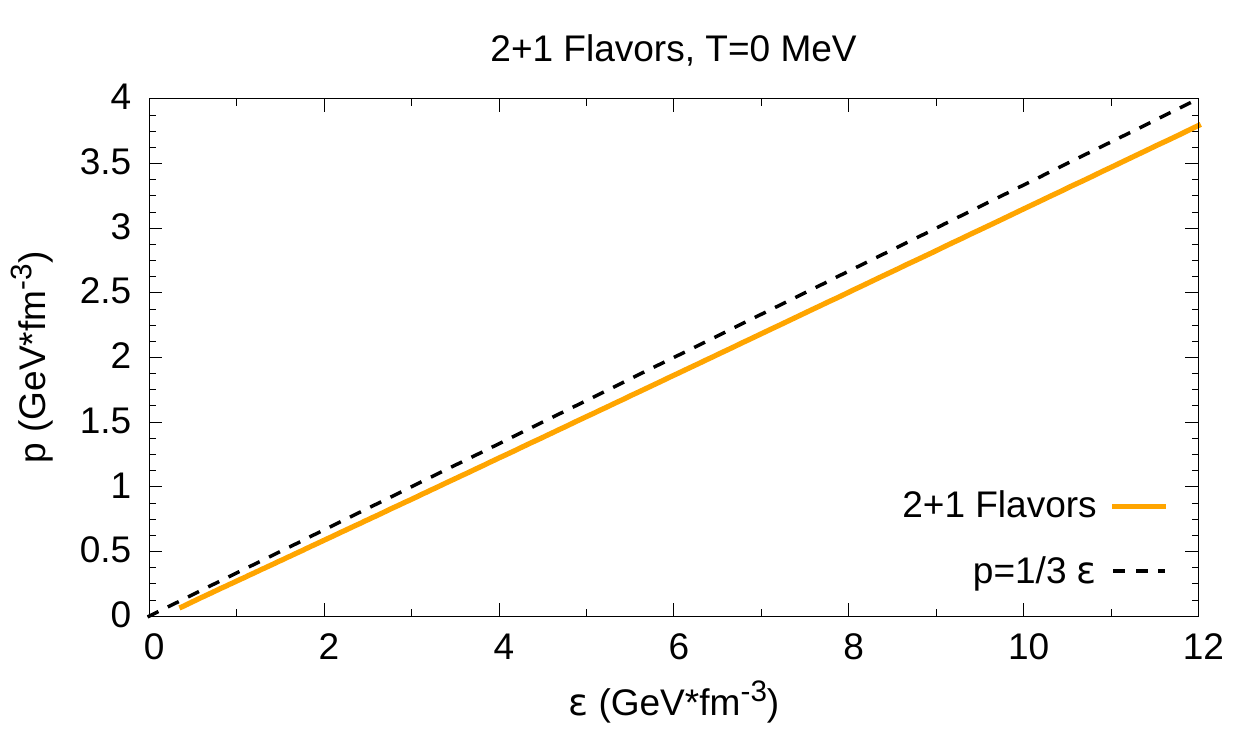}
    \end{minipage}
    \caption{Equation of state compared to the ideal case.}
    \label{fig:5}
  \end{figure}
\end{center}

Finally, in Fig.~\ref{fig:5} we plot the equation of state to exhibit the
deviations from ideality, i.e. $\varepsilon=3p$. The equation of state follows
closely the one for an ideal ultrarelativistic gas.

%%%%%%%%%%%%%%%%%%%%%%%%%%%%%%%%%%%%%%%%%%%%%%%%%%%%%%%%%%%%%%%
\section{Summary}

We investigated thermodynamic properties of massive cold quark matter at zero
temperature and baryon chemical potential and non-vanishing isospin density at
next-to-leading order in perturbation theory, and compared our results with
recent lattice data.

The ratio of energy density to (isospin density)$^{4/3}$ versus isospin
density shows that lattice data and our pQCD results get closer for high
densities. Both seem to follow a $\rho_{I}^{4/3}$ scaling at high densities,
which agrees with the limit for an ultra-relativistic degenerate Fermi gas. The isospin
chemical potential plotted against the isospin density shows that the pQCD
results and lattice results converge for values of $\mu_I\gtrsim3m_\pi$. This
is also true for the comparison of the normalized energy density as a function
of the isospin chemical potential. The normalized energy density is
essentially constant in the high-density limit, as expected.

We also verified that the energy density from the pQCD calculation is not too
far from the Stefan-Boltzmann limit for two flavors since the strange quark
does not appear in the dense medium under consideration. Furthermore, the 
deviations from an ideal equation of state are small.

In summary, the results from pQCD seem to be close to the lattice data already
for values of $\mu_I\gtrsim 3m_\pi$, even in the region of pion condensate. It
seems that the effect from the gap is suppressed for small values of the
coupling constant, as anticipated, and gives a small contribution to the
thermodynamic potential which is then dictated at high chemical potentials by
a nearly free gas of quarks.

%%%%%%%%%%%%%%%%%%%%%%%%%%%%%%%%%%%%%%%%%%%%%%%%%%%%%%%%%%%%%%%
\section*{Acknowledgments}

The authors want to thank Rainer Stiele, Lorenz von Smekal, Nils Strodthoff and William
Detmold for fruitful discussions. ESF is grateful for the kind hospitality of
the ITP group at Frankfurt University, where this work has been initiated. TG
is supported by the Helmholtz International Center for FAIR and the Helmholtz
Graduate School HGS-HIRe. The work of ESF is partially supported by CAPES,
CNPq and FAPERJ.

%%%%%%%%%%%%%%%%%%%%%%%%%%%%%%%%%%%%%%%%%%%%%%%%%%%%%%%%%%%%%%%
%\bibliographystyle{ieeetr}
%\bibliographystyle{apsrev4-1}
\bibliography{Paper_QCD_Isospin}

\end{document}